\def\rfr#1{eq. (\ref{#1})}

\def\virg#1{``#1''}

\def\eqi{\begin{equation}}
\def\eqf{\end{equation}}
\def\eqia{\begin{eqnarray}}
\def\eqfa{\end{eqnarray}}
\def\rp#1#2{{#1\over#2}} \def\lb#1{\label{#1}}

\def\bds#1{\boldsymbol{#1}}

\documentclass[11pt,twocolumn]{article}
\usepackage{amsmath,amsthm,amscd,amssymb,wasysym}
\usepackage{latexsym}
\usepackage{graphicx,epsfig}
\usepackage{abstract}
\RequirePackage{color}

\begin{document}

\twocolumn[
	
\noindent{\bf \LARGE{On some critical issues of the LAGEOS-based tests of the Lense-Thirring effect}}
\\
\\
\\
{L. Iorio$^{\ast}$}\\
{\it $^{\ast}$Ministero dell'Istruzione, dell'Universit\`{a} e della Ricerca (M.I.U.R.)\\ Fellow of the Royal Astronomical Society (F.R.A.S.)\\
 Permanent address: Viale Unit$\grave{a}$ di Italia 68
70125 Bari (BA), Italy.  \\ e-mail: lorenzo.iorio@libero.it}

\begin{onecolabstract}
We summarize some critical issues pertaining the tests of the general relativistic Lense-Thirring effect performed by I. Ciufolini and coworkers in the gravitational field of the Earth with the geodetic satellites LAGEOS and LAGEOS II tracked with the Satellite Laser Ranging technique.
\end{onecolabstract}

	]

\section{Introduction}
In the scalar theory of gravitation by Newton, which does not fulfil the Lorentz invariance, the gravitational field of a spherical body does not depend on its state of motion, being, indeed, determined only by its mass $M$. On the contrary, in the tensorial  General Theory of Relativity (GTR) by Einstein, which is a fully relativistic theory of gravitation, non-static distributions of matter-energy  yield their own contributions  to the overall gravitational field in addition to the static ones. Such peculiar terms are connected with the off-diagonal components $g_{0i}, i=1,2,3$ of the spacetime metric tensor, and were dubbed \virg{gravitomagnetic} \cite{Thorne,Rin,MashNOVA} in analogy with the magnetic fields generated by the electric currents.
Indeed, in its weak-field and slow-motion approximation, the fully non-linear field equations of GTR get linearized, thus resembling those of the linear Maxwellian electromagnetism; in particular\footnote{The quantities $\eta_{\mu\nu},\ \mu,\nu=0,1,2,3$ are the components of the metric tensor of the \virg{flat} Minkowskian spacetime.},  $g_{0i}\doteq \eta_{0i}+h_{0i},\ \left|h_{0i}\right|\ll 1,\ i=1,2,3$. The resulting gravitomagnetic part of the equations of motion of a test particle is \cite{Brum}
\eqi \ddot{x}^{i}=c\left(h_{0i,k}-h_{0k,i}\right)\dot x^k,\ i=1,2,3,\eqf where  dots denote  time derivatives, and $c$ is the speed of light in vacuum.

At great distances $r$ from a  slowly rotating body endowed with proper angular momentum $\boldsymbol{S}$,
it is
\eqi h_{0i}=-\rp{2}{c^2}A_i, i=1,2,3,\eqf
with the gravitomagnetic vector potential given by \cite{MashNOVA}
\eqi \bds A=\rp{G}{c}\rp{\bds S\bds\times\bds r}{r^3},\eqf
where $G$ is the Newtonian constant of gravitation.
The related gravitomagnetic field is \cite{Thorne0,Thorne,Mashetal01a}
\begin{equation}
\boldsymbol{B}_g(\boldsymbol{r}) \doteq \bds\nabla\bds\times\bds A= -\rp{G}{cr^3}\left[\boldsymbol{S} -3\left(\boldsymbol{S}\cdot\hat{r}\right)\hat{r} \right].\lb{gmfield}
\end{equation}
Among other phenomena, the resulting Lorentz-like, non-central acceleration due to \rfr{gmfield}
\eqi\bds A_{\rm LT}=-2\left(\rp{\bds v}{c}\right)\bds\times\bds B_g\eqf
causes the secular precession of the spin $\boldsymbol{\sigma}$ of a gyroscope (Schiff spin-spin effect, \cite{Sci}) with
frequency
\eqi \Psi_g = \rp{B_g}{c},\eqf and the
secular precessions of the longitude of the ascending node $\Omega$
\eqi\dot\Omega_{\rm LT}=\rp{2GS}{c^2 a^3(1-e^2)^{3/2}},\eqf
and the argument of pericenter $\omega$ \eqi\dot\omega_{\rm LT}=-\rp{6GS\cos I}{c^2 a^3(1-e^2)^{3/2}}\eqf
of the orbit of a test particle (Lense-Thirring spin-orbit effect, \cite{LT}),  both in geodesic motion around the spinning body. The parameters $a,e,I$ are the semimajor axis, the eccentricity and the inclination, respectively, of the particle's orbit. Such phenomena are often collectively denoted with the catchy denomination \virg{frame-dragging}, although also the general relativistic gravitoelectric  de Sitter  precession \cite{Des} of an orbiting gyroscope in the field of a static mass is a part of such a category \cite{Oco}.

Experimental/observational efforts have been dedicated in recent times to obtain empirical corroborations of the aforementioned predictions of GTR. The Gravity Probe B (GP-B) mission \cite{gpb1,gpb2} is an extremely refined, sophisticated and expensive experiment \cite{mad,bark,brum}, conceived 50 years ago, explicitly aimed to measure, among other things, the gravitomagnetic Schiff effect with four gyroscopes in a controlled environment enclosed in an active spacecraft  orbiting the spinning Earth since 2004. The properly scientific phase of the mission ended in 2005, and the analysis of the data collected during it is still ongoing \cite{gpb,gpb3,gpb4}. The expected accuracy was originally $1\%$ or better, but it seems that the occurrence of some unexpected systematic errors \cite{brum,gpb5,gpb6} may finally undermine the actual attainment of such a goal.  At present, according to the official mission's website\footnote{See http://einstein.stanford.edu/ on the WEB.}, the claimed statistical error is $14\%$, while the systematic uncertainty is $10\%$.

Attempts to measure the Lense-Thirring effect were proposed, and in some cases implemented, with some non-dedicated artificial and natural satellites in the Solar System; for a recent, comprehensive overview see, e.g., Ref.~\cite{megaiorio} and references therein. Concerning the performed analyses, the first tests date back to the mid of 90s \cite{ciuold1,ciuold2,ciuold3,ciuold4}; they were conducted in the gravitational field of the Earth with the non-dedicated LAGEOS and LAGEOS II geodynamic satellites\footnote{They are dense, spherical targets entirely covered with retroreflectors for passively bouncing back the laser impulses sent to them by  ground-based stations. Both the LAGEOS spacecraft orbit at altitudes of about 6000 km, so that they do not suffer macroscopic orbit decay due to the atmospheric drag. As a consequence, their lifetime is evaluated to be of the order of $10^5$ yr. } continuously tracked  with the Satellite Laser Ranging (SLR) technique \cite{SLR} by looking at the nodes of both the satellites and the perigee of LAGEOS II. Such attempts are still ongoing \cite{Ciu04,Ciu05,Ciu06,Ciu07,Ciu09,Ciu010} by retaining only the nodes of LAGEOS and LAGEOS II. The claimed accuracy in such more recent tests is $10-15\%$ \cite{Ciu04,Ciu05,Ciu06,Ciu07,Ciu09,Ciu010}, but other evaluations, dealing with certain sources of systematic errors in a more conservative way, point towards figures which may be up to $2-3$ times larger: for such critical views, see  Refs.~\cite{ioriossr,megaiorio} and references therein.

In this paper we want to clearly point out some epistemological and physical issues pertaining the performed LAGEOS-based analyses  which have not yet been explicitly addressed in a satisfactorily way. In Section \ref{test}, after a brief review of the status of the GP-B mission and the perspectives for performing other measurements of the Schiff precession with artificial and natural probes, we discuss if independent tests of the Lense-Thirring  effect really exist in literature after about 15 years since the first attempts were implemented. Section \ref{metodo} is devoted to the relation among the general relativistic gravitomagnetic field of the Earth and the spacecraft-based global  solutions for the classical part of the terrestrial gravitational field produced so far. Some alternative approaches to process the data of the LAGEOS and LAGEOS II spacecraft are discussed in Section \ref{lageosdata}. The issue of the actual level of cancelation of the corrupting bias due to the classical quadrupole mass moment of the rotating Earth in the tests performed so far is tackled in Section \ref{combs}; in it the impact of the other  multipoles of the terrestrial gravitational field according to the first models from GOCE is discussed as well. Section \ref{conclu} summarizes our findings.
\section{Do really independent  tests of frame-dragging exist?}\lb{test}
Physics is an activity whose results are considered as (provisionally) established if the experiments/observations which yielded them have been subsequently  repeated by different teams of independent researchers in different laboratories with different methodologies. Actually, this is not (yet ?) the case for gravitomagnetism.
\subsection{The Schiff effect and the GP-B experiment}
Concerning  GP-B, if, on the one hand, the analysis of the data collected in $2004-2005$ could, both in principle and in practice,  be repeated by other independent researchers, on the other hand it will likely not be possible to do that for the entire experiment in any foreseeable future in view of its extreme sophistication and cost.

This is certainly not satisfactorily from an epistemological point of view because GP-B seems destined to remain a unique empirical check of the Schiff prediction. Indeed, a proposal to use spacecraft orbiting the Sun and Jupiter \cite{haas} had not sequel so far; on the other hand, its complexity, cost and technological difficulties would certainly not have been lower than those of GP-B itself.

Moreover, independent measurements of the Schiff effect with natural bodies in, e.g., the Solar System are in all probability unfeasible. To this aim, let us recall that the maximum value of the gravitomagnetic Schiff precession occurs when the angular momentum $\boldsymbol {S}$ of the central source and the precessing spin $\boldsymbol{\sigma}$ of the gyroscope are mutually perpendicular, being, instead, zero when they are aligned \cite{Bark}. In principle, a natural scenario satisfying such a requirement is  the Sun-Uranus system. Indeed, while the solar equator is inclined to the mean ecliptic at the epoch J2000 by the Carrington angle $i_{\odot}=7.15$ deg \cite{carrin}, the spin $\boldsymbol{\sigma}_{\uranus}$ of Uranus is tilted to the ecliptic by $97.77$ deg \cite{ura}. Of course, apart from the difficulties of devising some effective methods for continuously monitoring the precessional motion of the spin of Uranus, the magnitude of the Sun-Uranus Schiff precession would be insignificantly small. In principle, another potential natural laboratory may, perhaps, be the double pulsar PSR J0737–3039A/B \cite{doppia,doppia2}. Indeed,  while the spin\footnote{Since the rotational periods of A and B are 23 ms and 2.8 s, respectively,  $\boldsymbol S_{\rm A}$  is larger than $\boldsymbol{\sigma}_{\rm B}$. The latter one describe a full precessional cycle in 75 yr because of the general relativistic de Sitter precession \cite{des,des2}.} $\boldsymbol{S}_{\rm A}$ of A is perpendicular to the orbital plane \cite{ferd}, $\boldsymbol{\sigma}_{\rm B}$ is not aligned with $\boldsymbol{S}_{\rm A}$ \cite{des,des2} because of the de Sitter precession \cite{Des} which has recently been measured with a $13\%$ accuracy \cite{des,des2}. Actually, the gravitomagnetic Schiff-like spin precession \cite{Bark} of $\boldsymbol{\sigma}_{\rm B}$  caused by $\boldsymbol{S}_{\rm A}$ would be much smaller and quite difficult to measure.
\subsection{The LAGEOS-based tests}
The situation is, in principle, more favorable for the Lense-Thirring tests performed with the LAGEOS satellites.

The first attempts to reveal the existence  of the Earth's gravitomagnetic field by analyzing the data of   LAGEOS and LAGEOS II  are $14-15$ years old, dating back to 1996-1997 \cite{ciuold1,ciuold2,ciuold3}.

The network of the\footnote{See http://ilrs.gsfc.nasa.gov/ on the WEB.} International Laser Ranging Service (ILRS) \cite{ilrs} consists of a large set  of laser ranging stations disseminated throughout the world, so that the SLR community is quite numerous. The LAGEOS satellites, which are some of the most important SLR targets, are continuously tracked  since long time. The GEODYN software \cite{GEODYN}, developed by NASA, is widely disseminated throughout the SLR stations also because it is free of charge. Moreover, some institutions developed their own orbit analysis systems like UTOPIA by the Center for Space Research (CSR) of the University of Texas, and the Earth Parameter and Orbit System (EPOS) by GeoForschugsZentrum (GFZ).

Despite this situation, potentially favorable for performing several truly independent tests of such a prediction of GTR, none has been actually either performed so far, or published in international peer-reviewed journals. Indeed, apart from a pair of conference talks given by J. Ries et al. (CSR) \cite{ries1,ries2}, all the relatively more accurate\footnote{One of the major critical points of the earlier tests was the use of the perigee of LAGEOS II \cite{ries3}, heavily perturbed by several non-gravitational classical forces.} tests published so far in peer-reviewed papers or edited books have I. Ciufolini in the authorship as first author or editor himself \cite{Ciu04,Ciu05,Ciu06,Ciu07,Ciu09,Ciu010}. Moreover, to date, no independent works on the Lense-Thirring effect attributable to members of GFZ  exist in literature. Thus, although the list of co-authors of the papers by Ciufolini has often changed and in some of the most recent works \cite{Ciu09,Ciu010} unpublished results obtained with UTOPIA and EPOS are described as well, such tests cannot be considered as truly independent ones.
This is particularly true also in view of the fact that the methodology adopted is basically the same, apart from the orbital processors used. This point will be explained better in Section \ref{metodo} and Section \ref{lageosdata}.

Thus, it will be possible to speak about genuinely independent tests of the Lense-Thirring effect with the LAGEOS satellites only if and when papers published in peer-reviewed journals without Ciufolini in the authorship, and authored if possible by  different researchers with respect to those having more or less systematically co-operated with him, will appear in literature. Moreover, and, perhaps, most importantly, also the methodology used should be different from that adopted so far.
Such studies, which should be made publicly available, would be of great importance even in the case of negative and/or inconclusive outcomes.
Finally, somebody may likely wonder why the author \textcolor{black}{of the present paper} does not undertake himself the task that he is suggesting to others. It may be pointed out that if, on the one hand, the results presented by skilful and experienced researchers in satellite data processing have raised doubts until now, on the other hand it is likely that analogous uncertainties would be even stronger in the case of a work produced by a scientist not yet actively engaged in such a difficult art. Moreover, the reliability of such results may, perhaps, be reduced in the eyes of a part of the community in view of the fact that the author \textcolor{black}{of the present paper} would difficultly be considered as sufficiently neutral and detached from the subject considered. Analyses by  really independent third parties may have more chances to be accepted without some sorts of prejudices.

Concerning the non-negligible role played by such considerations, it maybe instructive to illustrate the following case. In late 2007 a preprint titled \virg{A critical analysis of the GP-B mission. I: on the impossibility of a reliable measurement of the gravitomagnetic precession of the GP-B gyroscopes}, authored by G. Forst, was posted on the arXiv repository \cite{Forst}. This author never either posted other preprints on the arXiv website or published any peer-reviewed papers. Moreover, there is no mention at all on the WEB of the organization quoted as his affiliation. Finally, the references  cited by G. Forst did not actually show what was attributed to them in the main text of his preprint, as noted first by\footnote{See on the WEB http://www.physicsforums.com/show\-thread.php?t=104694$\&$page=18, post  $\#$282.} K. Krogh.
In early 2008, the arXiv moderators removed the preprint by G. Forst with the following motivation: \virg{This submission has been removed because 'G. Forst' is a pseudonym of a physicist based in Italy who is unwilling to submit articles under his own name. This is in explicit violation of arXiv policies.
Roughly similar content, contrasting the relative merits of the LAGEOS and GP-B measurements of the frame-dragging effect, can be found in pp. 43-45 of \cite{Ciu07}.} Even so, in late 2008 Ref.~\cite{Forst} was cited by I. Ciufolini in some talks of him \cite{forst1,forst2,forst3}.

At a different level of relevance, we also mention the\footnote{See http://en.wikipedia.org/wiki/Talk:Frame-dragging$\#$Recent$\_$controversy on the WEB.}
editing-war which involved the voice \virg{frame-dragging} on Wikipedia in 2006-2007.
It mainly consisted of the systematic and reiterated removal by Italian IPs followed by their almost immediate reinstatement, of all and only the references by the author \textcolor{black}{of the present paper} on some critical aspects of the Lense-Thirring tests with the LAGEOS satellites. On the contrary, the references by I. Ciufolini were never  removed.
\section{New global Earth's gravity field solutions}\lb{metodo}
A distinctive feature of all the global Earth's gravity field solutions \cite{models} obtained so far by several independent institutions from the data  of the dedicated spacecraft CHAMP \cite{CHAMP}, GRACE \cite{GRACE} and GOCE \cite{GOCE} is that GTR was never  explicitly solved for along with, say, the even zonal harmonic coefficients $\overline{C}_{\ell, 0},\ell=2,4,6,...$ of the geopotential. The first global solution, obtained from CHAMP,  dates back to 2001 \cite{champ}. Both the previous ones and all those of the CHAMP/GRACE/GOCE era are publicly available on the Internet at the official website of the International Centre for Global Earth Models (ICGEM), http://icgem.gfz-potsdam.de/ICGEM/.
This fact implies, among other things, that the even zonals may retain a sort of a-priori \virg{imprinting} of the Lense-Thirring effect itself; similar arguments were put forth in the pre-CHAMP/GRACE/GOCE era in Refs.~\cite{ries3,nord}. In Ref.~\cite{imprint} it was explicitly shown that this may actually be just the case for GRACE, given the present-day level of accuracy in estimating the even zonal coefficients and the size of the gravitomagnetic effect on the orbit of GRACE.
Although likely time consuming, producing new global Earth's gravity field solutions by explicitly solving for relativity as well would be a really important and independent test of the general relativistic gravitomagnetic component of the field of the Earth, also because it would allow to inspect the correlations among the estimated solve-for relativity parameter(s) and the even zonals in the covariance matrix \cite{ries3}. It would be important to judge if it made sense to employ that particular gravity model in processing the  data of the LAGEOS satellites to try to safely extract the Lense-Thirring effect.
The inquiries of the present author to some scientists presently involved in the generation of the global gravity field solutions have, in general, received rather evasive answers, if any, mainly centered on the issue of the great computational and time efforts which would be required to re-process all the data sets from, say, GRACE.
\section{A different approach in processing the LAGEOS data}\lb{lageosdata}
An issue related to the previous one is the approach followed to directly extract the Lense-Thirring effect from the data of the LAGEOS satellites, common to all the analyses performed so far. The directly observable quantity in SLR is the station-spacecraft range computed in terms of the two-way time-of-flight recorded by a ground-based clock. Actually, in all the tests implemented so far the gravitomagnetic effect on the LAGEOS ranges was never explicitly modelled in terms of one or more dedicated solve-for parameters to be estimated in the usually least-square sense \cite{nord}, as done, instead, for a host of other parameters pertaining certain physical properties of the spacecraft, their orbital motions and the Earth-fixed stations. Note that in the pre-CHAMP/GRACE/GOCE era the global Earth's gravity field solutions were produced just in such a way, i.e. by globally fitting long data records from a constellation of SLR targets, among which LAGEOS and LAGEOS II always played a dominant role, and estimating the even zonal harmonics as solve-for parameters. Incidentally, let us note that even in such circumstances the Lense-Thirring effect was never modelled and solved-for.

Another approach which may be followed may consist of not modeling the Lense-Thirring effect at all, and estimating in a purely phenomenological way some corrections $\Delta\dot\Omega$ to the node precessions. This may typically be done for each orbital arcs. A similar approach  was  followed in the determination of the corrections $\Delta\dot\varpi$ to the standard perihelion precessions of some planets of the Solar System \cite{Pit}, which, in principle, account for any unmodelled/mismodelled dynamical effect like just the Lense-Thirring one. Also in the case of  the timing of the binary systems hosting one or more pulsars a post-Keplerian  periastron precession $\dot\omega_{\rm PK}$ was  phenomenologically estimated as a solve-for parameter along with other ones \cite{pul}. Subsequently, it was  identified with the gravitoelectric precession predicted by GTR.

Future independent LAGEOS-based tests should try to implement such strategies. In doing that,   those global solutions which will be intentionally produced without retaining a-priori \virg{imprinting} of relativity itself should be used as reference gravity field models (see Section \ref{metodo}).
\section{Issues pertaining the bias due to the Newtonian multipoles of the terrestrial gravitational field}\lb{combs}
\subsection{The cancelation of the first even zonal harmonic of the geopotential}
All the most recent tests performed so far rely upon a linear combination of the nodes of LAGEOS and LAGEOS II  purposely designed to cancel out the  impact of the first even zonal harmonic coefficient $J_2\doteq -\sqrt{5}~\overline{C}_{2,0}$ of the multipolar expansion of the Newtonian gravitational potential of the Earth, which is a major source of systematic uncertainty. Indeed, it turns out that the nominal values of the competing secular node precessions of LAGEOS and LAGEOS II caused by $J_2$ are 7 orders of magnitude larger than the gravitomagnetic ones. Such a combination was  explicitly worked out in Ref.~\cite{combi}, following a strategy put forth in Ref.~\cite{Ciu96}; see also Refs.~\cite{ries3,ries4,pav}.

Actually, the coefficient $c_1$ of such a combination
is a function of the semimajor axes $a$, the eccentricities $e$ and the inclinations $I$ to the Earth's equator of characterizing the orbits of both the LAGEOS satellites.
This implies that the unavoidable uncertainties in the computation of such Keplerian orbital elements from the estimated state vectors of the satellites yield an overall uncertainty in $c_1$ itself which, thus, can be known only with a certain numbers of significant digits \cite{combierr}. In turn, this fact introduces a further source of systematic bias because, given a certain uncertainty $\delta c_1 \approx 1\times 10^{-8}-8\times 10^{-9}$  depending on the level of accuracy with which one assumes in determining the inclinations $I$ of LAGEOS and LAGEOS II, the resulting cancelation of the $J_2-$induced node precessions is necessarily not perfect, contrary to what implicitly assumed so far. It turns out that the residual $J_2$ signature would amount to $14-23\%$ of the Lense-Thirring one \cite{combierr}. It should be remarked that a value of $c_1$ known up to the $8-9$th decimal digits should be used to obtain just the aforementioned level of accuracy in the cancelation of the effect of $J_2$. Instead,  $c_1$ has always been treated so far  with a very limited number of decimal digits; in, e.g., Ref.~\cite{Ciu09} they are just 3 ($c_1=0.545$).
\subsection{The bias due to the other even zonal harmonics of higher degree}
Concerning the other even zonals of higher degree $J_{\ell}\doteq\textcolor{black}{-}\sqrt{2\ell +1}~\overline{C}_{\ell, 0},\ell=4,6,8,...$, which are not canceled out by the aforementioned linear combination,
their mismodeling induces a systematic uncertainty in the resulting signal which may amount to a non-negligible fraction of the Lense-Thirring signal. The realistic evaluation of such a systematic alias was evaluated in several papers; see, e.g., Refs.~\cite{ioriossr,megaiorio} and references therein.

Let us briefly recall that nowadays we have at our disposal several estimated values of the even zonals of the geopotential produced by different institutions with variable approaches and techniques from the data collected by the CHAMP, GRACE and GOCE dedicated missions. In evaluating the aliasing of the even zonals on the Lense-Thirring signal it should be clear that, since we are dealing with the same physical quantities simply measured with different techniques, there are  a-priori no reasons, in principle, to prefer just one specific solution instead of other ones, unless objective and quantitative arguments are provided for trusting just it. Certainly, in the framework of a test of fundamental physics, it is not acceptable to pick-up just this or that particular Earth'gravity models that, for some reasons, yield the best result in terms of fitted straight line\footnote{The issues previously discussed in Section \ref{metodo} and Section \ref{lageosdata} should, at this point, not be forgotten. }, and evaluating the systematic error on the Lense-Thirring effect by only using  such  particular solutions; it would be a sort of selection bias towards that solution just yielding  the closest outcome to the one expected in advance. Instead, it is much more realistic and conservative to take into account a large number of gravity models, provided that they are roughly of comparable accuracy, and adopt the differences among their estimated values for each even zonals as representative of the realistic uncertainty in them.
In any case, quantitative, statistical arguments should be used to discard one or more determinations of a given even zonal, as discussed in Ref.~\cite{megaiorio}. Stated simply, it is not admissible to play with the various gravity models by retaining only those convenient to the a-priori, desired outcome and discarding, instead, those yielding less favorable results.

In this respect, here we point out that the first,  preliminary global solutions from GOCE like\footnote{See http://portal.tugraz.at/portal/page/por\-tal/TU$\_$Graz/Einrichtungen/Institute/Homepa\-ges/i5080/forschung/GOCO/ on the Internet.} GOCO01S, which combines data from GRACE and GOCE, still present significative discrepancies with respect to earlier GRACE/CHAMP-only global solutions.
Indeed, it can be shown that the pair\footnote{They both use the tide-free system and the fully normalized norm, so that it makes sense to compare them.} GOCO01S-EIGEN51C, where EIGEN51C \cite{eigen51} is a global solution consisting of 6 years of CHAMP and GRACE data
and the DNSC08 global gravity anomaly data set, yields an uncertainty of $27\%$ of the Lense-Thirring signal. Another similar example is given by the pair GOCO01S-AIUB-GRACE02S yielding an uncertainty as large as $23\%$ of the Lense-Thirring signature; AIUB-GRACE02S \cite{aiub2} is a tide-free GRACE-only based solution obtained from almost 2 yr of GRACE data.
\section{Summary and conclusions}\lb{conclu}
Whatever the final outcome of its data analysis will be, the Gravity Probe B mission, explicitly dedicated to measure the general relativistic gravitomagnetic Schiff spin-spin effect in an extremely sophisticated and expensive controlled experiment carried onboard a spacecraft orbiting the Earth,
will  remain the only empirical check of this \textcolor{black}{specific} prediction of the General Theory of Relativity because of the practical impossibility of repeating it in any foreseeable future.
Moreover, no other natural laboratories in astronomical scenarios can likely be used to put on the test the  Schiff \textcolor{black}{gyroscope precession}.

In principle, the situation for the tests of the  Lense-Thirring spin-orbit effect performed so far in the gravitational field of the Earth with the non-dedicated LAGEOS and LAGEOS II satellites tracked with the Satellite Laser Ranging technique is more favorable. Indeed, the lifetime of such orbiting laser targets is of the order of $10^5$ yr, they are totally passive not requiring active instrumentation carried onboard, and the laser ranging community is made of several teams disseminated in a wide network of ground stations mainly using an orbital processor system which is freely available. Instead, despite the first attempts were made about 15 years ago by a group led by I. Ciufolini,  no really independent tests have been published so far in peer-reviewed journals by authors different from the aforementioned Italian scientist\textcolor{black}{, apart from a couple of conference talks by a group led by J. Ries}. On the contrary, fake Internet-based attempts to undermine the credibility of the Gravity Probe B mission \textcolor{black}{were} undertaken by an Italian physicist.

New Earth's global gravity field solutions in which the General Theory of Relativity is explicitly solved for along with the multipolar coefficients of the classical part of the geopotential should be produced. On the contrary, all the models obtained so far from the dedicated CHAMP, GRACE and GOCE missions by several independent institutions since 2001 may be a-priori \virg{imprinted} by the General Theory of Relativity \textcolor{black}{itself} since no relativistic effects \textcolor{black}{were} ever  explicitly estimated in them.

A closer connection between the gravitomagnetic effects on the orbits of LAGEOS and LAGEOS II  and the directly observable quantities in Satellite Laser Ranging should be elucidated. In this regard,  the Lense-Thirring effect should be explicitly modeled in the dynamical force models of the LAGEOS satellites\textcolor{black}{,} and a dedicated parameter should be solved-for, as it is common practice in all other areas of space science and gravitational physics.

The cancelation of the first even zonal harmonic coefficient of the classical multipolar expansion of the terrestrial gravitational potential, which is of degree 2, from the linear combination of the nodes of LAGEOS and LAGEOS II used so far is not perfect because of the uncertainties in their orbital parameters. It turns out that the uncanceled effect of the Earth's centrifugal oblateness is as large as $14-23\%$ of the Lense-Thirring combined signal.

 Significative discrepancies among the estimated values of the even zonal harmonics in the first, preliminary models from GOCE and the earlier models from CHAMP and GRACE exist; according to them, the systematic uncertainty caused by the mismodeling in the even zonals of degree higher than 2 on the Lense-Thirring signature of LAGEOS and LAGEOS II is still as large as about $20\%$.
 \section*{Acknowledgements}
 I thank  M. Cerdonio for insightful and inspiring correspondence \textcolor{black}{occurred in September 2010}.


\end{document}